\documentstyle[aps,multicol,epsf]{revtex}

\newcommand{\be}{\begin{equation}}
\newcommand{\ee}{\end{equation}}

\begin{document}
\preprint{SNUTP 98-132}
\draft
\widetext

\title{Possibility of direct Mott insulator-to-superfluid transitions\\
       in weakly disordered boson systems} 
\author {Sung Yong Park$^{1}$, Ji-Woo Lee$^{1}$, Min-Chul Cha$^{2}$,
         M.Y. Choi$^{1}$, B.J. Kim$^{1,*}$, and Doochul Kim$^{1}$} 
\address {$^{1}$Department of Physics and Center for Theoretical Physics, 
Seoul National University\\
Seoul 151-742, Korea\\ 
$^{2}$Department of Physics, Hanyang University, Kyunggi-do, Ansan 425-791,
Korea}

\maketitle

\thispagestyle{empty}

\begin{abstract}
We study the zero-temperature phase transitions of a two-dimensional
disordered boson Hubbard model at incommensurate boson densities.
Via matrix diagonalization and quantum Monte Carlo simulations,
we construct the phase diagram and evaluate the
correlation length exponent $\nu$.
In the presence of weak disorder,
we obtain $\nu=0.5 \pm 0.1$, the same value as that in the pure model,
near the tip of a Mott insulator lobe, using the dynamical critical exponent $z=2$.
As the strength of disorder is increased beyond a certain value, 
however, the value of $\nu$ is found to change to $0.9 \pm 0.1$.
This result strongly suggests that
there exist direct Mott insulator-to-superfluid transitions
around the tip of a Mott insulator lobe in the weak disorder regime.
\end{abstract} 

          
\begin{multicols}{2}
\narrowtext 

Two-dimensional (2D) interacting boson systems display
quantum phase transitions~\cite{Sondhi} from an insulating state to
a superfluid (SF) state at zero temperature.
Physical realization of this transition
may include disordered thin film superconductors, Josephson-junction arrays,
granular superconductors~\cite{superconductor},
and $^4$He films adsorbed in porous media~\cite{porous}.
In the absence of disorder the insulating state is a Mott insulator (MI), 
which has a commensurate value of the boson density
since, otherwise, excessive bosons or holes
will move freely to yield a superfluid.
The excessive bosons or holes can be localized, on the other hand,
in the presence of disorder,
to make an incommensurate insulator.
The resulting Bose glass (BG) insulating phase has attracted
considerable interest recently. 
In the mean-field theory~\cite{Fisher},
it has been argued that in the presence of disorder
the transition from a Mott insulator to 
a superfluid occurs only through the BG phase.
However, a recent quantum Monte Carlo study~\cite{Rieger}, 
performed at the tip of an MI lobe (i.e., at a commensurate value of the
boson density), has demonstrated the existence of a direct MI-SF transition.
The subsequent renormalization group study~\cite{Zimanyi}
has been interpreted to suggest that the direct MI-SF transition occurs 
around the tip of the MI lobe
in high dimensions ($d>4$) but only at the tip in lower dimensions ($2\leq d<4$). 
These results raise an interesting question as to whether 
such a direct MI-SF transition occurs even at incommensurate values of the
boson density.

This work investigates the possibility of the direct MI-SF transition
off the tip of an MI lobe in two dimensions.
We perform both matrix diagonalization and quantum Monte Carlo simulations,
and find evidences for the direct 
transition around the tip of an MI lobe in the weak disorder limit.
The superfluid onset points are identified and the correlation
length critical exponents are estimated at various disorder strengths.
The corresponding phase diagram for weak disorder is constructed.

We consider the boson Hubbard model with disorder,
described by the Hamiltonian
\begin{eqnarray}
H=\frac{U}{2}\sum_i n_i^2-\sum_i(\mu+v_i) n_i
-t\sum_{\langle i,j\rangle} (b_i^{\dag} b_j + b_i b_j^{\dag}),
\label{eq:H1}
\end{eqnarray}
where $b_i^{\dag}$ and $b_i$ are the boson creation and
destruction operators at site $i$ on an $L\times L$ square lattice, and
$n_i \equiv b_i^{\dag} b_i$ is the number operator.
In Eq.~(\ref{eq:H1}), $U$ is the strength of the on-site repulsion,
$\mu$ is the chemical potential,
$v_i$ is the random on-site potential
distributed uniformly between $-\Delta$ and $\Delta$,
and $t$ measures the hopping strength between
nearest neighboring sites.
In the limit of the large number of particles, 
we may take the phase-only approximation~\cite{phase only} 
and reduce Eq.~(\ref{eq:H1}) to the quantum phase Hamiltonian 
\begin{eqnarray}
H=\frac{U}{2}\sum_i n_i^2-\sum_i(\mu+v_i)n_i
  -t\sum_{\langle i,j\rangle} \cos(\phi_i-\phi_j),
\label{eq:H2}
\end{eqnarray}
where $\phi_i$ is the phase of the bosons condensed at site $i$
and satisfies the relation $\left[n_i, \phi_j\right]=i\delta_{ij}$.
Note that in this representation, $n_i$ denotes the deviation 
from the mean integer number $n_0$ 
(around the mean number $\langle b_i^{\dag} b_i\rangle$).

In order to determine the phase boundary,
we first use the matrix diagonalization method with a truncated basis set.
The zero-temperature phase boundary between the MI
phase and the SF/BG phase is determined by 
comparing the ground state energy of the system at commensurate boson density
and that of the system containing an extra hole or particle.
The basis set is chosen to include the lowest-energy states
of the Hamiltonian given by the first and the second term in Eq.~(\ref{eq:H2})
(i.e., the zeroth order in $t$) and
the states coupled to these by hopping up to $(2n{+}1)$th order in $t$,
where we set $n=1$ in this work.
We identify the lowest-energy state among the states of which the
total boson number is zero as the MI state and the lowest one among the states
with the total boson number being unity as the SF/BG state, 
and determine the phase boundary by locating the values of $\mu\,(>0)$ and $t$
at which the energies of the two states become equal.

Since the basis set includes those states coupled up to the 3rd order in $t$,
the result will be the same as that from the energy perturbation 
to the same order.  The numbers of states involved in 
this calculation are $4L^4{-}9L^2$ and $4L^2+1$ in the cases of the BG/SF state 
and of the MI state, respectively. 
We adopt the Lanczos method~\cite{Lanczos} to obtain the lowest eigenvalue 
and the corresponding eigenstate, and
take 2000 disorder realizations for each system size.  
In the pure case ($\Delta=0$), we find the phase diagram which is consistent with the 
previous perturbation result of Freericks and Monien~\cite{Monien}.

\begin{figure}
\hskip -0.4cm
\epsfxsize=8cm \epsfysize=6cm \epsfbox{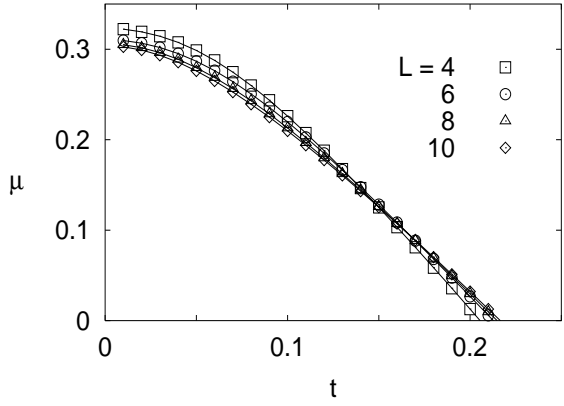}
\caption{Phase diagram in the $(t, \mu)$ plane 
for the disorder strength $\Delta=0.2$. 
Each line gives an estimate of the phase boundary between
the MI state and the SF or BG state, as 
determined by matrix diagonalization.}
\label{fig:phase}
\end{figure}

Figure \ref{fig:phase} shows the zero temperature phase diagram with the 
disorder strength $\Delta=0.2$ up to the system size $L = 10$
obtained from the truncated basis set.
The shape of the MI lobe constructed from the previous perturbation 
result in the pure case~\cite{Monien}
should be round and shorter than that obtained here.
In the absence of hopping ($t = 0$), where no SF state is possible, 
the phase boundary (between the MI and BG states)
approaches, as the system size $L$ is increased, $0.5{-}\Delta$, as it should.

To distinguish the SF phase from the BG phase, we use localization argument,
and define the participation ratio as
\begin{equation}
p_L=\left[\frac{\left(\sum_i p_i^2\right)^2}{\sum_i p_i^4}\right]_{av},
\end{equation}
where $p_i$ is the probability that particles are found at site 
$i$~\cite{Rieger}, and $[\cdots]_{av}$ denotes the average over different
disorder realizations.
At the generic SF-BG transition, the participation ratio
satisfies the scaling relation
\begin{equation}
\frac{p_L}{L^2}=L^{-y} \tilde{p}\left(\delta_t L^{1/\nu}\right),
\end{equation}
where $\delta_t=(t-t_c)/t_c$ is the distance from the critical point $t_c$ and
$\nu$ is the correlation length exponent. Here the scaling function $\tilde{p}$
and an additional exponent $y$ have been introduced~\cite{Rieger,Scalettar}. Note
that since we just consider the states in which the total number of bosons is unity,
this method is useful to determine the critical point very close to the MI phase,
i.e., on the phase boundary in Fig.~\ref{fig:phase}.

\begin{figure}
\hskip -0.4cm
\epsfxsize=8cm \epsfysize=6cm \epsfbox{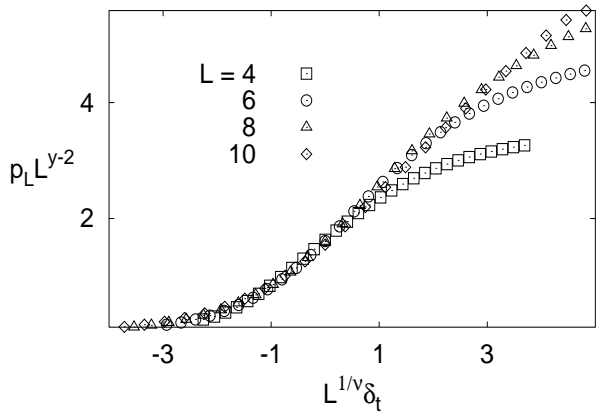}
\caption{Scaling plot of the participation ratio $p_L$ near the MI-SF/BG 
phase boundary, which yields $t_c = 0.12$, $y = 0.92$, and $\nu = 1.5$.}
\label{fig:scaling}
\end{figure}

Figure \ref{fig:scaling} presents the finite-size scaling of the participation 
ratio of the SF/BG state.
From this scaling behavior,
we obtain $y=0.92 \pm 0.03$, $\nu=1.5 \pm 0.3$, and $t_c=0.12 \pm 0.02$;
the latter
separates the SF state from the BG one
as $t$ (or $\mu$) is varied along the phase boundary in Fig.~\ref{fig:phase}. 
Accordingly, the system undergoes a direct MI-SF transition 
as the phase boundary is crossed for small $\mu$ ($\lesssim 0.18$ in Fig.~\ref{fig:phase}), 
i.e., near the tip of an MI lobe.
Note that the possible origin of this direct MI-SF transition, as discussed later, is the
abundance of particle-hole excitations around the tip of a lobe.
This suggests that the size of the basis set including the
states up to the 3rd order in $t$ might not be sufficient to determine
the phase boundary unambiguously and to compute the corresponding exponents accurately.
Such abundance of particle-hole excitations 
is expected to forbid the perturbation calculation to be accurate,
possibly causing the discrepancy between the perturbative results~\cite{Monien} 
and the quantum Monte Carlo results~\cite{Otterlo} in the pure case without disorder.

It is thus needed to investigate the transition via 
quantum Monte Carlo simulations, which are in general more reliable.
For that purpose,
we follow the standard procedure~\cite{Otterlo,Young} to transform
the 2D quantum phase Hamiltonian in Eq.~(\ref{eq:H2}) to the 
(2+1)-dimensional classical action
\begin{equation}
S =\frac{1}{K}\sum_{(i,t)}^{\nabla\cdot J=0} 
\left[\frac{1}{2} {\bf J}^{ 2}_{(i,t)} -(\mu+v_i)J^{\tau}_{(i,t)}\right],
\label{eq:H3}
\end{equation}
where the integer current vector
${\bf J}_{(i,t)}=(J^x_{(i,t)}, J^y_{(i,t)},J^{\tau}_{(i,t)})$
is divergenceless on each lattice
site $(i,t)$ as indicated.
The coupling constant $K$, corresponding roughly to $\sqrt{t/U}$,
takes the role of the temperature.

We perform quantum Monte Carlo simulations on the classical action in Eq.~(\ref{eq:H3}),
employing the heat bath algorithm~\cite{Young} at a classical temperature $K$.
An important quantity in the analysis is the zero-frequency
superfluid stiffness
\begin{equation}
\rho=\frac{1}{L_\tau}\left[\left\langle n_x^2\right\rangle\right]_{av}, 
\end{equation}
where $n_x=(1/L)\sum_{(i,t)} J^x_{(i,t)}$ is the winding number
along the (spatial) $x$ direction.
The finite-size scaling behavior of the superfluid stiffness~\cite{Young,Cha}
reads
\begin{equation}
\rho= L^{-(d+z-2)} \tilde{\rho}\left(L^{1/\nu} \delta,L_\tau/L^z\right),
\end{equation}
where $\delta=(K{-}K_c)/K_c$ is the distance from the critical point $K_c$,
the spatial dimension $d$ is two in this work,
and $z$ is the dynamical critical exponent.

In order to investigate the scaling behavior,
the value of $z$ should be known in advance.
At the very tip of an MI lobe, 
the direct MI-SF transition in the presence of weak disorder
shows the same behavior as the pure system:
the dynamical critical exponent $z=1$
and the correlation exponent $\nu=0.67$~\cite{Rieger}.
Off the tip, we expect the dynamical exponent $z = 2$
in the disordered system since the compressibility is finite at the
transition point~\cite{Fisher}; the same number is expected
even in the pure case.
We thus set the dynamical exponent $z = 2$, and
measure the correlation length exponent $\nu$,
the value of which was estimated in the previous
studies to be $0.9\pm 0.1$ in the BG-SF 
transition~\cite{Rieger,Young} and $0.5 \pm 0.1$ in the MI-SF 
transition~\cite{Fisher,Otterlo}.

Keeping $L_\tau/L^z$ constant,
we simulate the systems of sizes
$L\times L \times L_\tau = 6\times 6\times9$, $8\times 8\times 16$, and 
$10\times 10\times 25$.
To tune the transition, we vary the temperature $K$ while fixing $\mu$ and $\Delta$.
We further take the average over $60-200$ disorder realizations
and, for each disorder realization, perform typically $4000-80000$
Monte Carlo sweeps for equilibration,
followed by equally many sweeps for measurement.
The equilibration is checked through the use of
the standard equilibration test technique~\cite{Young,Bhatt}.

\begin{figure}
\hskip -0.4cm
\epsfxsize=8cm \epsfysize=6cm \epsfbox{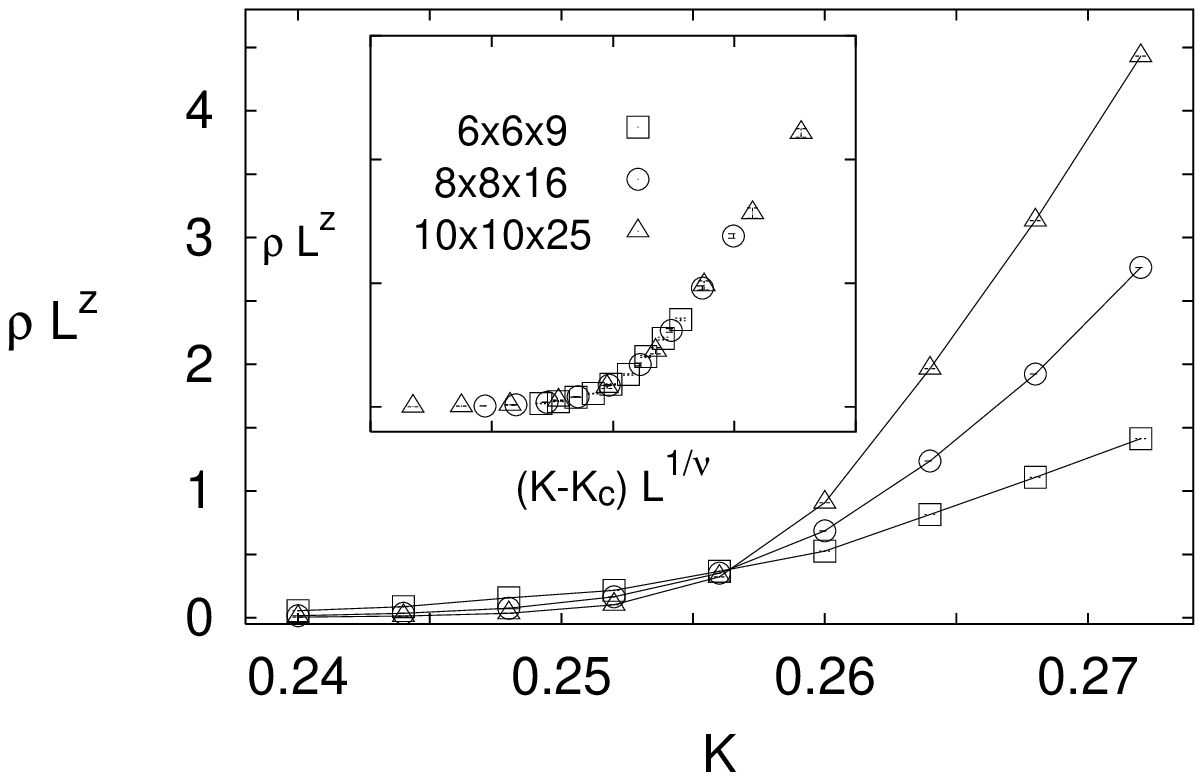}
\caption{
Behavior of the stiffness $\rho$ for $\mu=0.3$ and $\Delta=0.1$,
with the aspect ratio kept constant for $z = 2$.
The inset shows a scaling plot with $K_c = 0.257$ and $\nu=0.5$.
}
\label{fig:weak}
\end{figure}

\begin{figure}
\hskip -0.4cm
\epsfxsize=8cm \epsfysize=6cm \epsfbox{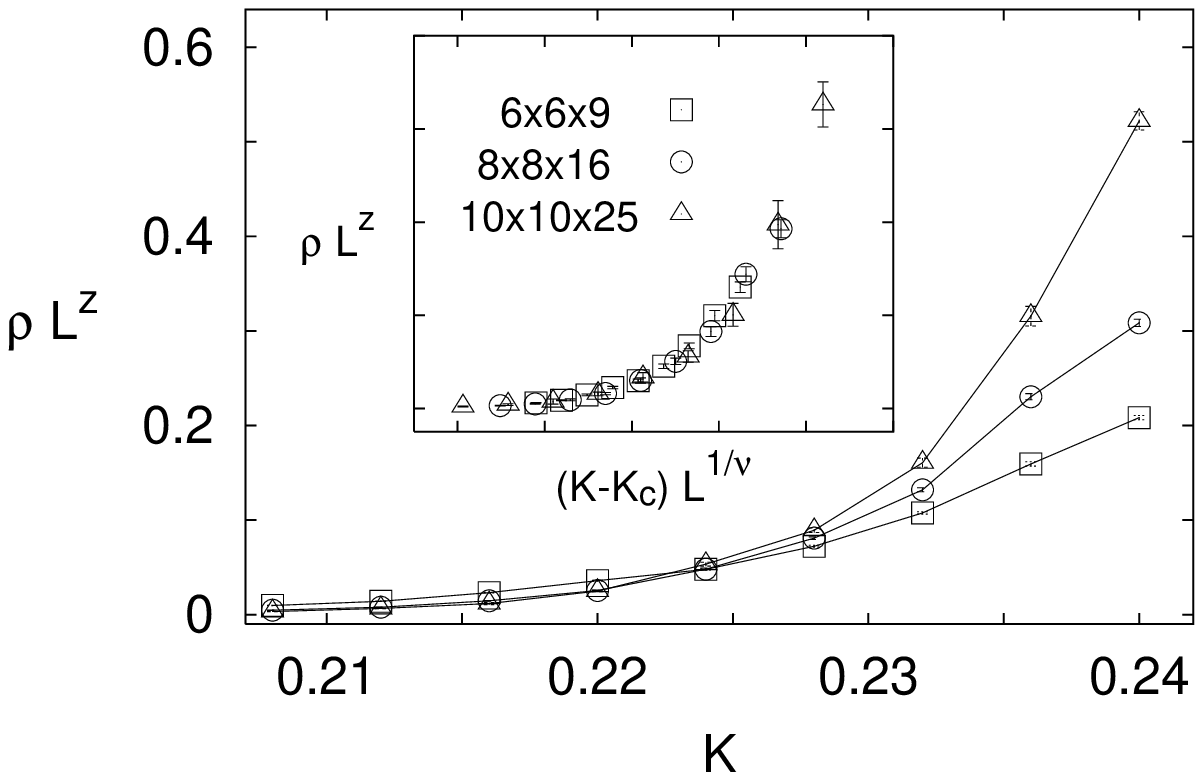}
\caption{
Behavior of the stiffness $\rho$ for $\mu=0.3$ and $\Delta=0.3$,
with the aspect ratio kept constant for $z = 2$.
The inset shows a scaling plot with $K_c = 0.223$ and $\nu=0.9$.
}
\label{fig:strong}
\end{figure}

\begin{figure}
\hskip -0.4cm
\epsfxsize=8cm \epsfysize=6cm \epsfbox{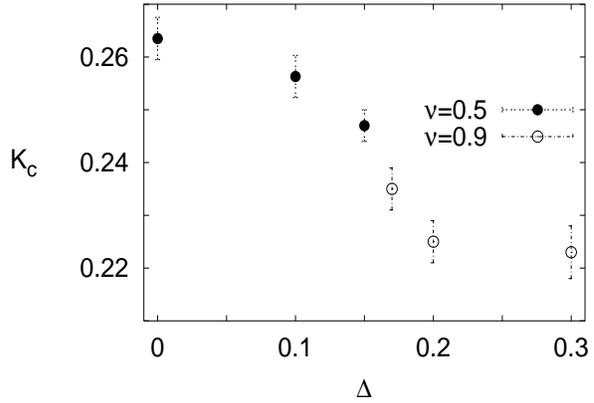}
\caption{Critical point $K_c$ as a function of the disorder strength
$\Delta$ for $\mu = 0.3$. The solid dots indicate the direct 
MI-SF transition points and the open dots the BG-SF transition points.
}
\label{fig:disorder}
\end{figure}

In Fig.~\ref{fig:weak}, we show the results with $z=2$ at $\mu=0.3$
and $\Delta=0.1$.
One can identify clearly the common crossing point 
of $L^z\rho$ at $K_c=0.257 \pm 0.003$ as the critical point. 
The inset of Fig.~\ref{fig:weak} 
shows a scaling plot, which yields $\nu=0.5\pm 0.1$,
the same value as the pure model~\cite{Fisher}.
On the other hand, in the strong-disorder case ($\Delta=0.3$) shown in 
Fig.~\ref{fig:strong}, the scaling behavior
yields $\nu=0.9\pm 0.1$, which agrees well with the previous 
results of the BG-SF transition~\cite{Fisher,Rieger,Young}.
These demonstrate that in the weak disorder regime
a direct MI-SF transition takes place not only at the tip 
but also off the tip of the MI lobe.
Figure \ref{fig:disorder} presents the critical temperature
$K_c$ as a function of the disorder strength $\Delta$ for $\mu = 0.3$,
displaying the direct MI-SF transition occurring for $\Delta < 0.16$.
It is of interest that the point on which the value of
the correlation length exponent $\nu$ changes appears
to have the maximum slope on the $K_c${-}$\Delta$ curve.

\begin{figure}
\hskip -0.4cm
\epsfxsize=8cm \epsfysize=6cm \epsfbox{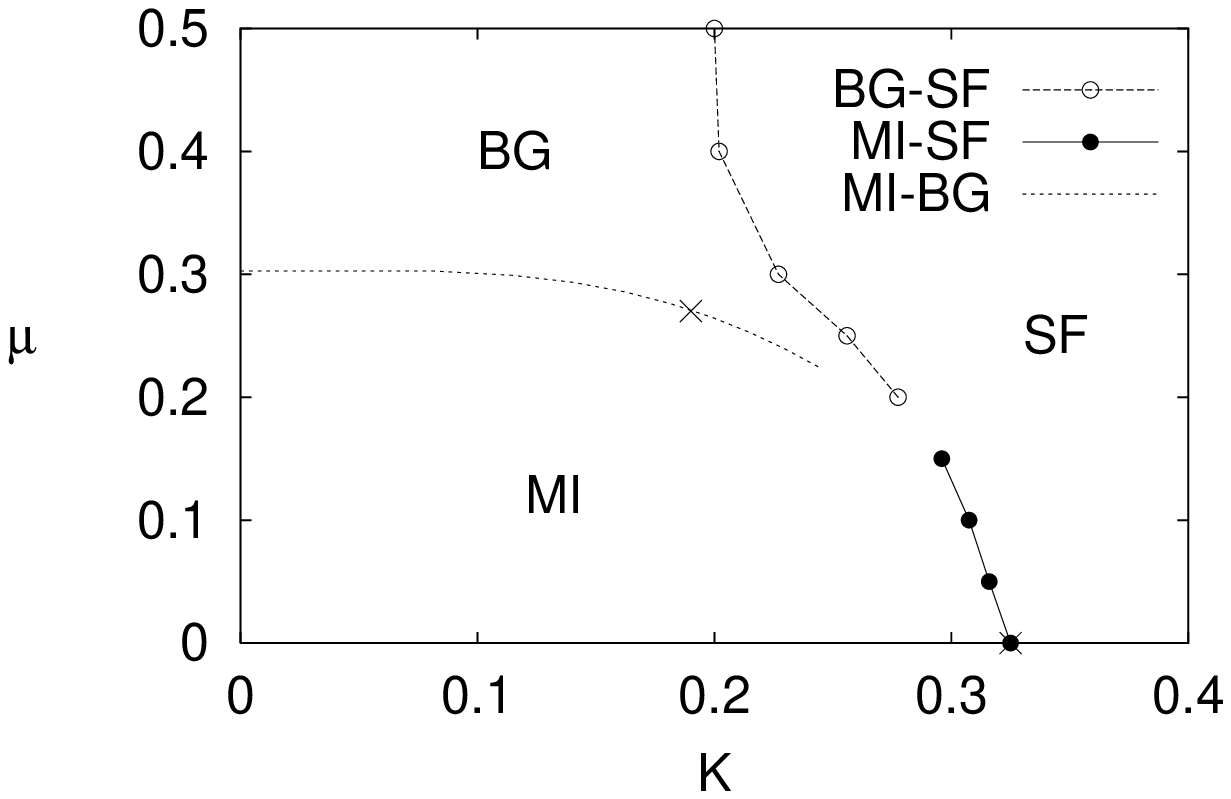}
\caption{
Phase diagram in the $(K, \mu)$ plane for $\Delta = 0.2$.
The dotted line (without symbols) corresponds to the rescaled data
from the matrix diagonalization method, while the solid and dashed
lines (with symbols) are the results of quantum Monte Carlo simulations. 
The two points marked by crosses represent the results in Ref.~[5].
}
\label{fig:phase_diagram}
\end{figure}

Figure~\ref{fig:phase_diagram} summarizes the phase diagram of the 
disordered boson Hubbard model at the disorder strength $\Delta = 0.2$.
The phase boundary between the MI and the BG phases is estimated
by means of the matrix diagonalization in the system of size $L = 10$.
Here we have used the results in Ref.~\cite{Rieger,Monien} to scale $K$ as a
function of $t$, which is valid only for small $t$.

Finally, we discuss the possible origin of the direct MI-SF transition around 
the tip of an MI lobe. 
One might think that off the tip the BG phase intervenes very slimly 
between the MI and SF phases and that
our results supporting the direct MI-SF transition merely reflect finite-size effects.
Be this the case, the localization length would be large near the tip and
the BG phase would manifest itself only on the length scale exceeding the system size.
Then, as pointed out in Ref.~\cite{Zimanyi}, an anomalous behavior is expected to occur 
to change the correlation exponent $\nu$,
suggesting continuous change of the exponent
with the thickness of the BG phase near the tip.
On the other hand, our results in Figs. \ref{fig:disorder} and 
\ref{fig:phase_diagram} indicate that the exponent changes quite abruptly, 
making the above scenario rather unlikely.
Near the tip, the particle-hole excitations are ubiquitous and
presumably tend to suppress the disorder effects, 
allowing the possibility of an extended state for the extra boson.
Thus the resulting direct MI-SF transition around the tip may reflect the peculiar 
nature of boson localization.

In summary, we have studied the disordered boson Hubbard model 
by means of the matrix diagonalization with restricted basis states,
which include those states overlapping with each other through nearest-neighbor
hopping up to the 3rd order.
The finite-size scaling of the
participation ratio at $\Delta = 0.2$ gives evidence for the direct MI-SF
transition at an incommensurate value of the boson density.
To investigate this direct transition more clearly,
we have investigated the scaling behavior of
the superfluid stiffness via quantum Monte Carlo simulations
and found that as the disorder strength is varied, 
the value of the correlation exponent $\nu$ 
changes rather abruptly from the weak-disorder value $0.5$ to the strong-disorder
one $0.9$.
This indicates that the direct MI-SF 
transition occurs for weak disorder at an incommensurate density.
The possible origin of this transition could
be the abundance of the particle-hole excitations around the tip of an MI lobe,
which suggests the peculiar nature of boson localization.

S.Y.P. would like to thank H. Rieger, G. G. Batrouni, and J. Kisker for helpful 
discussions.  This work was supported in part by the Ministry of Education 
through the BSRI Program and by the KOSEF through the SRC Program.
The work of M.C.C. was also supported in part by the BSRI Program through
Hanyang University.

\end{multicols}

\end{document}